\documentclass[prd,superscriptaddress,amsfonts,amssymb,amsmath,showpacs,twocolumn,showkeys]{revtex4-1}
\usepackage{amsmath}
\usepackage{epsfig}
\usepackage{graphics}
\usepackage[colorinlistoftodos]{todonotes}
\usepackage{color}
\usepackage{graphicx}
\usepackage[font={footnotesize,it}]{caption}
\usepackage[colorlinks=true,
            linkcolor=black,
            urlcolor=blue,
            citecolor=blue]{hyperref}
    
\def\be{\begin{equation}} 
\def\ee{\end{equation}}
\def\beq{\begin{eqnarray}}
\def\eeq{\end{eqnarray}}
\def\n{\nonumber}

\begin{document}

\title{Equilibrium stellar configurations in Rastall theory and linear equation of state}

\author{Sudan Hansraj}
\email{hansrajs@ukzn.ac.za}
\affiliation{Astrophysics and Cosmology Research Unit, University of KwaZulu Natal, Private Bag X54001, Durban 4000,
South Africa.}

\author{Ayan Banerjee}
\email{ayan_7575@yahoo.co.in}
\affiliation{Astrophysics and Cosmology Research Unit, University of KwaZulu Natal, Private Bag X54001, Durban 4000,
South Africa.}

\date{\today}

\begin{abstract}

Amongst a number of modified  theories of gravity, the Rastall theory has been propounded to address some shortcomings of the standard theory of general relativity. Our purpose is to investigate this framework's capacity to analyse  stellar structure in the context of elementary requirements for physical plausibility such as positive definite functions for the energy density and pressure, conformity to the causality criterion and the existence of an equation of state. We consider the analogue of the Saslaw \textit{et al} \cite{saslaw} isothermal model of general relativity and show that the Rastall version satisfies the basic requirements unlike its counterpart. Then we examine in turn the consequences of suppressing one of the inverse square law fall off of the energy density or the linear equation of state. In addition, the case of a constant spatial gravitational potential is studied on account of this prescription being a necessary and sufficient condition for isothermal behaviour in Einstein theory and its most general tensor extension Lovelock gravity.

\end{abstract}

%\keywords{}

\maketitle

\section{Introduction}
\label{sec:intro}

It is generally believed that realistic astrophysical models are expected to display an equation of state (EoS) relating the internal pressure with the energy density. Typically stars manifest either a polytropic equation of state in the form of a power law expression or a simplistic linear barotropic equation of state. The analysis of stellar structure with these properties have routinely followed the numerical path in view of the intransigence of the associated Einstein field equations in yielding an exact solution with a prescribed equation of state. Numerical codes suffer the drawback of admitting errors which are compounded when implemented repeatedly in the model. An exact solution does not suffer this drawback and the evolution of models may be studied with greater accuracy.

As explained, the field equations of general relativity are difficult to solve if an EoS is imposed at the outset. For the configuration of a static fluid sphere some 120 exact solutions \cite{delgaty,finch} have emerged with the overwhelming majority obtained through the avenue of specifying some geometric or dynamical prescriptions. Later the model is investigated to check if an EoS is admitted. One notable exception is the isothermal fluid model of Saslaw {\it{et al}} \cite{saslaw} where an inverse square falls off of the density is required {\it{ab initio}} as well as a linear equation of state.  It is well known that for this case of matter the number of independent Einstein equations are three, while there are four variables to determine, namely, the two gravitational potentials, the density ($\rho$) and pressure ($p$). Therefore, the choice of two prescriptions by \cite{saslaw} appears to over-determine the system and one of the equations, the pressure isotropy equation is used as a consistency condition and so all the field equations are satisfied. Some of the solutions derived by Tolman \cite{Tolman} also display an EoS, but these emerged at the end of the solution generating process. Still there is no known general exact solution for the Einstein field equations for a static perfect fluid sphere with the equation of state  $p = \eta \rho$ for some constant $\eta$. Numerical studies were accomplished by Nilsson and Uggla for polytropes in \cite{nilsson1} and and for linear
equations of states in \cite{Nilsson2}. 

In this work we examine the role of the linear barotropic equation of state in the context of the Rastall theory of gravity  \cite{Rastall,Rastall1}. 
In  this modification to  Einstein's theory 
the usual law of the conservation of the energy momentum tensor does not hold.  The initial motivation of this modification is to challenge the well known conservation of energy-momentum  in the curved spacetime without violating the Bianchi identities. Within this paradigm  the covariant divergence of energy-momentum tensor is proportional to the covariant divergence of the curvature scalar, i.e., $T^{\mu}_{\nu;\mu}$ $\propto$ $R_{;\nu}$.
As a result, the non-minimal coupling leads to a modified general relativity theory.  The main point in favor of the Rastall theory is that in a flat  spacetime or as a  first approximation of a weak gravitational field, one can recover all the known laws of conservation. Another feature of Rastall theory  is the action principle or matter field Lagrangian is not yet decidedly known and work on this problem is ongoing. An important advance was made by dos Santos and Nogale \cite{dossantos} who generated a Lagrangian density for Rastall gravity that seemed to indicate that Rastall gravity is a special case of $f(R,T)$ gravity theory. Visser \cite{Visser} recently erred in claiming that Rastall theory was equivalent to Einstein gravity \cite{Hansraj,Darabi}. 

To mitigate the manifest shortcomings in terms of the non-traditional view of energy conservation evident in Rastall theory, it is noted that similar structures like those of  Rastall's theory may be found in the context of Weyl geometry \cite{Almeida}. The merit of this theory with respect to others is related to the fact that  the field equations are simpler than those of other modified theories. Therefore it is now believed that this gravity model could lead  to some major differences in several problems of current interest, such as the well known problems related to current scenarios of the universe. Such phenomenological ideas are often propounded to correct anomalies in the standard theory such as the inability of general relativity to explain the current epoch of expanding acceleration the universe is undergoing.   

Some interesting physically important  results have been obtained within this theory.  In the framework of Rastall gravity, cosmological work has been preformed by Batista \emph{et al.} \cite{Batista1}. Based on this gravity theory (see \cite{Heydarzade,Kumar,Heydarzade1} and references therein), rotating  and non-black hole solutions have been investigated. In addition neutron stars (NS) have also been considered by Oliveira \emph{et al} \cite{Oliveira}. Thermodynamical properties 
 for both a static spherically symmetric metric and also in the flat FLRW universe have been studied \cite{Moradpour,Moradpour1}. Recently, in \cite{Hansraj} studied all Tolman solutions for perfect fluid sphere in the context of Rastall theory. 

Our interest lies in the implications of Rastall's theory to stellar structure development. Specifically we analyze the role of the linear equation of state. The analysis is aided by the presence of a certain Rastall parameter which allows for the detection of exact solutions more readily when compared to general relativity. In particular we follow the Saslaw {\it{et al}} \cite{saslaw} programme to determine the unique isothermal fluid solution in Rastall gravity. Next we check the consequences of removing the equation of state requirement but insisting on an inverse square law fall off of the density. This is sufficient to find a unique solution if it exists. In addition we study the situation where  only the linear equation of state is imposed and endeavour to locate exact solutions. It turns out that the most general solution is indeed intractable as expected, however for particular stipulations of the constants available, exact solutions are indeed obtainable. We study the viability of such solutions in representing realistic astrophysical phenomena by referring to elementary requirements such as the positive-definiteness of the density and pressure as well as the adiabatic stability indicated by the sound speed index. Additionally we compute the expressions indicating the bevahiour of the weak, strong and dominant energy conditions. 

 The paper is organized as follows.  After an introduction in section~\ref{sec:intro}, we briefly
 recall the basic construction of Rastall theory~\ref{spacetime}. Subsequently,
we derive the field equations for a spherically symmetric metric perfect fluid in section ~\ref{field}. Next we discuss the isothermal property in~\ref{isothermal}. It is shown that  the Einstein case violates the elementary physical requirements whereas the Rastall  isothermal model does not suffer these defects,
 In section \ref{Inverse}, the inverse square law fall-off of the density has been investigated in detail for a  particular choice of Rastall parameter. In section \ref{potential}, we assume a constant spatial metric potential 
and study the basic properties of stellar structure. The study is enhanced by considering graphical plots of a typical case.  In Section \ref{barotropic} we impose a linear barotropic EoS to close off the system of field equations and examine the consequences. Finally in section~ \ref{Con}, we conclude with a brief 
discussion of our results.

\vspace{0,5cm}

\section{ Brief review of Rastall theory of gravity}
\label{spacetime}

Let us review now the basic elements of the  theory of gravity proposed by  P. Rastall \cite{Rastall,Rastall1}. The  basic assumption is the fact that $T^{ab}_{;b}\neq 0$, i.e.,
the usual conservation law of the energy momentum tensor does not hold. The 
covariant divergence of the energy-momentum tensor is proportional to the 
covariant divergence of the curvature scalar. In particular, it can be written as 
\begin{equation}
T^{\mu\nu}_{;\mu} = \alpha R^{;\nu}, \label{1}
\end{equation}
where $R$ is the Ricci scalar, and $\alpha$ is the Rastall parameter 
which quantifies the deviation from the  Einstein theory of General Relativity (GR).
In GR, it is postulated that $T^{\mu\nu}_{;\mu} =0$, which means Rastall gravity
turns out to be a modification of Einstein's tensor where a non-minimal coupling of 
matter fields to geometry is considered. The modified Einstein tensor can 
be written as
\begin{eqnarray}
G_{\mu\nu}+ \gamma g_{\mu\nu}R= \kappa Tg_{\mu\nu}, \label{2}
\end{eqnarray}
where $\gamma = k \alpha$ and $k$ is the Rastall gravitational coupling
constant. In summary,  one can express the above equation in a compactified form as 
\begin{eqnarray}
G_{\mu\nu}= \kappa T_{\mu\nu}^{\text{eff}}, \label{3}
\end{eqnarray}
where $T_{\mu\nu}^{\text{eff}}$ represents the effective energy-momentum tensor, which is defined as
\begin{eqnarray} \label{4}
T_{\mu\nu}^{\text{eff}}= T_{\mu\nu}-\frac{\gamma T}{4\gamma -1}g_{\mu\nu}.
\end{eqnarray}
The expression for $T_{\mu\nu}^{\text{eff}}$ is given by \cite{Moradpour1}
\begin{eqnarray}
S^0_0 \equiv -\rho^{\text{eff}}  =-\frac{(3\gamma-1)\rho+\gamma(p_r+2p_t)}{4\gamma-1},\\\label{5}
S^1_1 \equiv p^{\text{eff}}_r  = \frac{(3\gamma-1)p_r+\gamma(\rho-2p_t)}{4\gamma-1},\\\label{6}
S^2_2= S^3_3\equiv p^{\text{eff}}_t  = \frac{(2\gamma-1)p_t+\gamma(\rho-p_r)}{4\gamma-1},\label{7}
\end{eqnarray}
where $\rho$ is the energy density, $p_r$ and  $p_t$ are the radial and tangential pressures, 
respectively which are in general different ($p_r \neq  p_t$) to allow for anisotropy. It is to be noted that the energy-momentum 
tensor is conserved when $\alpha \rightarrow 0$ as in the case of general relativity.
Also, for a traceless energy-momentum source, such as the electromagnetic source, the Eq.~(\ref{3}), leads 
to $T_{\mu\nu}^{\text{eff}}=T_{\mu\nu}$, and it benefits from the fact that standard 
Einstein gravity is again recovered. According to Rastall theory \cite{Rastall} the Eq. (\ref{2}) leads to $R(4 k \gamma-1)=T $, and 
this demonstrates  that the trace $T$ of the energy momentum tensor is not always zero. Therefore, the $\kappa \gamma=1/4$  
case is prohibited in this theory. Instead if  we consider the Newtonian limit and define the Rastall
dimensionless parameter $\gamma= \kappa \alpha$, then we have the following relations for  the coupling constant 
($\kappa$) and the Rastall parameter ($\alpha$) \cite{Moradpour1}
\begin{equation}
\kappa=\frac{4\gamma-1}{6\gamma-1}8\pi, ~~\text{and}~~\alpha=\frac{\gamma(6\gamma-1}{(4\gamma-1)8\pi}.\label{eq8}
\end{equation}
From the above relations we see that when $\alpha=0$, the Einstein result $\kappa =8\pi$ is regained  which is parallel to the $\gamma= 0$ limit \cite{Heydarzade}. On the other hand, when $\gamma=1/6$, the Rastall gravitational coupling constant diverges. 
Thus, we should also exclude the case of $\gamma=1/6$ in Rastall gravity. The final form of the Rastall's field equations is 
\begin{eqnarray}
G^{\mu}_{\nu}+ \gamma g^{\mu}_{\nu}R= 8\pi \frac{4\gamma-1}{6\gamma-1} T^{\mu}_{\nu}, 
\end{eqnarray}
which leads to $R(6 \kappa \gamma-1)= 8 \pi T $. This lead to $\gamma = \kappa \alpha = 1/6$ not allowed in this theory in agreement with Eq. (\ref{eq8}). In particular, for the values of $\gamma =1/6$ and $\gamma =1/4$, the Rastall theory 
does not make any physical sense.

\section{FIELD EQUATIONS}
\label{field}

The geometry we are interested here is static  spherical symmetry throughout this paper. 
The metric signature convention is taken to be $(-,+,+,+),$ with Greek indices running over spacetime coordinates. The metric is conveniently written in Schwarzschild-like coordinates $(t, r, \theta, \phi)$, with the line element
\begin{equation}
ds^{2} = -e^{\nu(r) } \, dt^{2}+e^{\lambda(r)} dr^{2} +r^{2}(d\theta ^{2} +\sin ^{2} \theta \, d\phi ^{2}), \label{met}
\end{equation}
where the gravitational potentials $\nu$ and $\lambda$ depend only on the
radial coordinate $r$. We consider the source is a perfect fluid distribution, which is
characterized by energy density $\rho(r)$ and isotropic pressure $p(r)$.
In the considered spacetime the energy-momentum tensor for a spherical distribution of matter 
is given by $T_{\mu\nu} = (\rho + p)u_{\mu} u_{\nu} + p g_{\mu\nu}$, where the Greek indices 
$\mu$ and $\nu$ run from 0 to 3, with $u_{\mu}$ is the fluid's four velocity.The choice of a
perfect fluid implies that flow of matter is adiabatic, no radiation, heat flow  or 
viscosity is present \cite{Misner}.

Now, using the space time metric given in Eq.~(\ref{met}) and the
non-vanishing trace part of the effective energy-momentum tensor into Eq.~(\ref{4}), 
the  equations of motion then reduce to the following set of three coupled ordinary 
differential equations (see for instance, Ref. \cite{Hansraj}) as 
\begin{eqnarray}
\frac{(4\alpha - 1)e^{-\lambda}}{r^2} \left(1 - r\lambda' + e^{\lambda}\right) = - 3\alpha p - (3\alpha -1)\rho , \n \\  \label{6a} \\
\frac{(4\alpha - 1)e^{-\lambda}}{r^2} \left(1 + r\nu' - e^{\lambda}\right) = (\alpha - 1) p + \alpha \rho ,\n \\  \label{6b}\\
r^2(2\nu'' + \nu'^2   - \nu'\lambda' ) - 2r(\nu' +\lambda')  + 4(e^{\lambda} - 1) = 0. \n \\\label{6c}
\end{eqnarray}
where the prime (') denotes differentiation with respect to, r. Note that
the Eq.~(\ref{6c}) is the pressure isotropy equation and identical to that of standard Einstein theory. The first two equations (\ref{6a})) and (\ref{6b}) can be expressed independently as
\begin{eqnarray}
\rho &=& \frac{e^{-\lambda}}{r^2 } \left(-\lambda_{1} - (\alpha -1)r \lambda' + 3\alpha r \nu'  \right), \label{7a}  \\
p &=& \frac{e^{-\lambda}}{r^2 } \left(\lambda_{1}
+ \alpha r \lambda' - (3\alpha - 1)r \nu'   \right). \label{7b} 
\end{eqnarray}
where we have introduce a new variable $\lambda_{1}=(4\alpha - 1)(e^{\lambda} - 1)$.  Combining the above equations one can obtain the inertial mass density, which is
\begin{eqnarray} \rho  +  p = \frac{e^{-\lambda} (\nu' + \lambda')}{r},
\end{eqnarray}
which is independent of the Rastall parameter, $\alpha$. 
Hence, we are left with a system of three differential
equations, (\ref{6a}), (\ref{6b}) and (\ref{6c}), which are not enough to solve for four variables $\lambda$, $\nu$, $\rho$, and $p$. It is required that one of the variables must be specified at the start and the remaining three determined through integrating the field equations. Alternatively, a functional dependence of one variable on another may be postulated  motivated on physical grounds. For example an EoS $p \equiv p(\rho)$ is understood to characterize realistic stellar distributions. 

One immediate observation from the Rastall field equations is that unlike in GR, the $G^t_t = T^t_t$ equation contains the density, pressure and potential variable $\lambda$. The consequence of this is that locating the analogue of Schwarzschild's interior metric in this framework is impeded and to date the exact solution for the incompressible fluid sphere in Rastall gravity is not known. In the standard theory the $G^t_t = T^t_t$ field equation contains only $\lambda$ and $\rho$. This means that choosing a functional form for $\lambda$ is equivalent to specifying the energy density and vice-versa. Hence setting $\rho = $ a constant, immediately fixes $\lambda$ and the isotropy equation may be invoked to determine $\nu$ for the Schwarzschild interior solution. This is not so straightforward in the Rastall framework and considerably more work is needed to untangle the differential equations.

\section{Exact solution with isothermal behaviour}
\label{isothermal}

Isothermal fluid is characterized by an inverse square law fall off of the energy density $\rho$ as well as the equation of state $p = \eta \rho$ for a constant $\eta$ such that $0 < \eta < 0$.  Furthermore introducing the isothermality property of $\rho \sim 1/r^{2}$, one can write the energy density and pressure in the following form $\rho =\frac{B}{r^2}$ and $p = \frac{A}{r^2}$ for some constants $A$ and $B$.
Inserting these functional forms in the field equations (\ref{7a}) and (\ref{7b}) generates the system 
\begin{eqnarray}
 Be^{\lambda} &=& (4\alpha -1)(1-e^{\lambda}) - (\alpha - 1)r \lambda' + 3\alpha r \nu{'} ,\label{8a} \\
 Ae^{\lambda} &=& (4\alpha - 1)(e^{\lambda} - 1) + \alpha r\lambda{'} - (3\alpha - 1)r\nu{'}. \label{8b}
\end{eqnarray}

Now, adding (\ref{8a}) and (\ref{8b}) we obtain the following relationship
\begin{eqnarray}
 \nu' =
\frac{(A+B)e^{\lambda}}{r} - \lambda', \label{9}
\end{eqnarray}
 expressing $\nu$ in terms of $\lambda$. Substituting (\ref{9}) in
 (\ref{8a}) generates the differential equation
\begin{eqnarray}
(4\alpha - 1)\left(1 - e^{\lambda} - r \lambda'  \right) - (B -
(3\alpha (A + B)) = 0. \n \\ \label{10}
 \end{eqnarray}
containing only $\lambda$ and is solvable as 
 \begin{eqnarray}
e^{\lambda} = \frac{(1-4 \alpha )^2 r {C_1}^{4 \alpha }}{C_1-(4
\alpha -1) r {C_1}^{4 \alpha  } \xi}, \n \\ 
\label{11}
\end{eqnarray}
where $C_1$ is a constant of integration. Inserting (\ref{11}) into
(\ref{9}) gives
\begin{eqnarray} e^{\nu} =\frac{1}{r} \left(C_1 -(4 \alpha -1) r
{C_1}^{4 \alpha } \xi\right){}^{\frac{-\alpha
(A+B+4)+A+1}{\xi}} + C_2, \n \\ \label{11b}
\end{eqnarray}
where $C_2$ is a new constant of integration. For notational simplicity we introduce $\xi =\alpha  (3 A+3 B-4)-B+1$. It now remains to determine the constants $A$ and $B$ such that
the pressure isotropy is satisfied. The consistency condition Eq. (\ref{6c})  reduces to 
\begin{eqnarray}
&&(4 \alpha -1) r {C_1}^{4 \alpha  } \left(A^2-4 \alpha
\left((8 A-1) B+5 (A-1) A+3 B^2\right)\right. \n  \\
&& \left.
+4 \alpha ^2 (A+B) (7 A+7
B-4)+6 A B-4 A+B^2\right) \n  \\ && + C_1 (-8 \alpha  (A+B)+5 A+B) = 0.  \label{12}
\end{eqnarray} 
 is identically satisfied provided that
\begin{eqnarray}
\{(A,B)\} = \{ (0; 0), (-1 + 8\alpha, 5 - 8\alpha)\}.
\end{eqnarray}

Hence, we have obtained an isothermal fluid sphere in the Rastall
framework with dynamical quantities
\begin{eqnarray} p = \frac{-1+8\alpha}{r^2}
\hspace{0.5cm} {\mbox{and}} \hspace{0.5cm} \rho =
\frac{5-8\alpha}{r^2}, \end{eqnarray}
where the trivial $(0; 0)$ solution  has been omitted. 
To ensure a positive definite density and pressure we require $\alpha$ to lie in the window $\left(\frac{1}{8}, \frac{5}{8} \right)$, while the velocity of sound speed is given by \begin{eqnarray} \frac{dp}{d\rho} =
\frac{-1+8\alpha}{5-8\alpha}.
\end{eqnarray}
A subluminal sound speed is
guaranteed in the interval $\alpha \in \left(\frac{1}{8},
\frac{3}{8}\right)$. In summary a causal and well behaved isothermal
fluid exists provided that $ \frac{1}{8} < \alpha < \frac{3}{8}$. Thus, the geometry is uniquely determined by the potentials
\beq
e^{\lambda} &=&  \frac{(1-4\alpha)^2 r}{C_1^{1-4\alpha} - 4(1-4\alpha)^2 r}, \label{12a} \\ \n \\
e^{\nu} &=& \frac{1}{r}+C_2,  \label{12b}
\eeq
which does not reduce to the Saslaw {\it{et al}} \cite{saslaw} metric previously found for standard Einstein gravity.

It should be noted that when $\alpha \rightarrow 0$ i.e., specializing to the Einstein case, the
pressure and density have  the simple forms $p = -\frac{1}{r^2} $ and $\rho = \frac{5}{r^2}$, while the
sound speed is given by $\frac{dp}{d\rho} = -\frac{1}{5}$. Clearly in the absence of the Rastall parameter
the corresponding Einstein model completely violates elementary physical requirements essential for stable stellar configurations at equilibrium. Interesting the  Saslaw {\it{et al}} \cite{saslaw} model also suffers this defect as it can be shown that no value of the proportionality parameter $\alpha$ (not to be confused with the Rastall parameter $\alpha$ used in this work) obtained from the equation of state $p = \alpha \rho$ exists such that the metric potentials $e^{\nu}$, $e^{\lambda }$, density and pressure are simultaneously positive and in addition satisfy the causality criterion also. Accordingly such a model is defective while our model above does indeed satisfy the basic requirements for physical admissibility provided that the Rastall parameter is nonzero. 

\bigskip

\section{Inverse square law fall-off of density}
\label{Inverse}

In this section, we analyse the consequence of abandoning the requirement of an EoS initially but requiring an inverse square law fall-off of the density. The mathematical problem is then well posed consisting of three equations with three unknowns so that theoretically a unique solution exists.  Let $\rho = \frac{B}{r^2}$ in (\ref{7a}), and then we introduce the  transformation $e^{\lambda} = b(r)$. The pressure isotropy equation then assumes the following form
\beq
b^2 \left(28 \alpha ^2-20 \alpha +(5 \alpha -2) r (4 \alpha +B-1) b'+1\right) &&  \n \\
+\left(-44 \alpha ^2+28 \alpha +(2-20 \alpha ) B-2\right) b^3  && \n \\
+(4 \alpha +B-1)^2 b^4+\left(-8 \alpha ^2+7 \alpha +1\right) r^2 b'^2 && \n \\  +  r b \left(6 (\alpha -1) \alpha  r b'' +\left(-20 \alpha ^2+13 \alpha -2\right) b'\right)  && = 0, \n \\ \label{41}
\eeq
which is nonlinear. It is difficult to obtain the general solution of Eq (\ref{41}), however special cases may yield exact solutions.

Here, we consider the case $\alpha = 1$ which is motivated by the vanishing of the $b''$ and $r^2 b'^2$ terms.  Then, Eq (\ref{41}) reduces to 
\be 3 b \left((B+3) r b'+3\right)+(B+3)^2
b^3-18 (B+1) b^2-9 r b' = 0 \label{42}, \ee
which is first order and solvable by quadratures. Rearranging the terms of (\ref{42}), one can write the expression as
\be \frac{b'(3(B+3)b - 9)}{b(b-k_1)(b-k_2)}
= \frac{1}{r}, \label{43} 
\ee
where we have defined $k_1 =
\frac{9(B+1) + 6\sqrt{B(B+3)}}{(B+3)^2}$ and $k_2 =
\frac{9(B+1) - 6\sqrt{B(B+3)}}{(B+3)^2}$.

Invoking partial fractions, the integration of (\ref{43}) may be accomplished in the implicit form 
\be b^{k_4}(b- k_1)^{k_3} (b-k_2)^{k_5} = Cr, \label{45} 
\ee 
where $C$ is an integration constant and we have put $k_3 = \frac{3(3(k_1
-1)+Bk_1)}{k_1(k_1 -k_2)}$, $k_4 = -\frac{9}{k_1 k_2} =
-\left(\frac{3}{B+3}\right)^2$ and $k_5 = -\frac{3(3(k_2
-1)+Bk_2)}{k_2 (k_1-k_2)}$. Observe that Eq. (\ref{45}) is an algebraic equation but  nontrivial to solve explicitly. We seek values of $B$ such that a solvable equation emerges. Scrutinizing the discriminant in $k_1$ and $k_2$ suggests the fortuitous value $B = 3$ whence $k_3 = 9$, $k_4 = -18$ and $k_5=9$. Also, $k_1 = 1 + \frac{\sqrt{2}}{2}$ and $k_2 = 1 - \frac{\sqrt{2}}{2}$.  Then Eq. (\ref{45}) assumes the form
\be
(b-k_1)(b-k_2) = \left(Cr \right)^{\frac{1}{9}} b^2, \label{46}
\ee
which is quadratic, and hence solvable as
\begin{eqnarray}
b = e^{\lambda} =  \frac{\pm  \sqrt{2(f+1)}-2}{2 (f-1)}, \label{47}
\end{eqnarray}
where we have redefined $f = (Cr)^{1/9}$.
Consequently the remaining gravitational potential is given by
\begin{eqnarray}
e^{\nu} = \frac{f^9 \left(\frac{1+\sqrt{f+1}}{1-\sqrt{f+1}}\right)^{9 \sqrt{2}} \left( \sqrt{2(f+1)}-2\right)^{18}}{\left(f-1\right)^{18} \left( \sqrt{2(f+1)}+2\right)^{18}}. \label{47a}
\end{eqnarray}
with the help of  Eq. (\ref{8b}). 
 Now, we are in a position to determine the energy density, pressure and the sound speed, respectively. Plugging the values of the metric potentials into the relevant equations we obtain
 \begin{widetext}
 \beq
 \rho &=&  \frac{3}{r^2} ,   \label{47b} \\ \n \\
 p &=&  -\frac{10 \sqrt{2} f+17  \sqrt{f+1} f-36 \sqrt{f+1}+9 \sqrt{2}}{9 r^2 \left(\sqrt{2} f-2 \sqrt{f+1}+\sqrt{2}\right)},   \label{47c} \\ \n \\
 \frac{dp}{d\rho} &=& \frac{-595 \sqrt{2} f^3 +3396 \sqrt{2} f +859 \sqrt{2} f^2 + (436f^2 -2800f -3240)  \sqrt{f+1}  +1944 \sqrt{2}}{972 \sqrt{f+1} \left(\sqrt{2} f-2 \sqrt{f +1}+\sqrt{2}\right)^2}.  \label{47d}
  \eeq
  \end{widetext}
 The EoS may easily be obtained by substituting $r^2 = \frac{3}{\rho}$ into the pressure function. 
  %Fig 1 demonstrates the qualitative   behaviour of the density (thick curve), pressure %(thin  curve) and sound speed index (dashed curve).
  The expressions governing the energy conditions evaluate to
  \beq
  \rho - p &=& \frac{37 \sqrt{2} f+(17f -90) \sqrt{f+1} +36 \sqrt{2}}{9 r^2 \left(\sqrt{2} f -2 \sqrt{f+1}+\sqrt{2}\right)},  \label{47e} \\ \n \\
  \rho + p & = &  \frac{\left(17 f +18\right) \left(\sqrt{2}-\sqrt{f+1}\right)}{9 r^2 \left(\sqrt{2} f-2 \sqrt{f+1}+\sqrt{2}\right)} ,   \label{47f} \\ \n \\
  \rho + 3p &=&  -\frac{\sqrt{2} f+(17f -18) \sqrt{f+1} f}{3 r^2 \left(\sqrt{2} f-2 \sqrt{f+1}+\sqrt{2}\right)}.  \label{47g}
  \eeq
 Recall that we perform all the calculations for the specific permissible Rastall parameter value  $\alpha = 1$. Evidently, the inverse square law fall-off of the density does not lead to a linear barotropic EoS although a functional dependence of pressure on density $p = \eta \rho$  explicitly exists. Notwithstanding the numerator of the pressure function vanishes only for $f = 1$ however the same is true for the denominator. Consequently no surface of vanishing pressure exists. Additionally, the pressure, sound speed, strong energy condition and dominant energy expressions are all negative. While such violations of the energy conditions may be characteristic of a dark energy model,  we do not analyze this model further. \\

\section{Constant spatial potential}
\label{potential}

A constant spatial gravitational potential is known to generate a fluid model with isothermal behaviour in the standard theory as well as its generalisation the pure Lovelock theory \cite{dad-hans}. We investigate the consequences of this prescription in the Rastall framework and we  set $e^{\lambda} = k$, for some constant $k$. As a result, the pressure isotropy Eq. (\ref{6c}) is solved to give 
\be
e^{\nu} = {c_2} r^{2-2 \sqrt{2-k}} \left(c_1+r^{2 \sqrt{2-k}}\right){}^2,  \label{50}
\ee
which is the temporal gravitational potential. The dynamical quantities now evaluate to
\beq
\rho &=& \frac{\alpha  \left(\frac{12 k_1 r^{2 k_1}}{c_1+r^{2 k_1}}-6 k_1 +10\right) +k(1-4 \alpha)  -1}{k r^2},   \label{51a} \\ \n \\
p &=&  \frac{(4 \alpha -1) (k-1)-\frac{2 (3 \alpha -1) \left(c_1(1- k_1) +\left(k_1+1\right) r^{2 k_1}\right)}{c_1+r^{2 k_1}}}{k r^2}, \n \\  \label{51b}
\eeq
where we have put $k_1 = \sqrt{2-k}$. It can be observed that isothermal behaviour arises in the special case $c_1 = 0$. In addition, the speed sound is given by
\begin{widetext}
\be
\frac{dp}{d\rho} = \frac{c_1^2 \left(k_3 +2 k_1-3\right)+2 (2 \alpha -1) c_1 (k-1) r^{2 k_1}+\left(k_4-2 k_1-3\right) r^{4 k_1}}{c_1^2 \left(1 -k_3\right)-2 (2 \alpha +1) c_1 (k-1) r^{2 k_1}+\left(1-k_4\right) r^{4 k_1}},  \label{52}
\ee
\end{widetext}
after relabeling $k_3 = (1-4 \alpha)  k -2\alpha  \left(3 k_1 -5\right)$ and $k_4 = (1-4 \alpha)  k+2 \alpha  \left(3 k_1+5\right)$. The expressions for  energy conditions evaluate to 
\begin{widetext}
\beq
\rho - p &=& \frac{2 \left(c_1 \left(10 \alpha -4 \alpha  k-6 \alpha  k_1 +k+k_1-2\right)+\left(10 \alpha -4 \alpha  k+6 \alpha  k_1+k-k_1-2\right) r^{2 k_1}\right)}{k r^2 \left(c_1+r^{2 k_1}\right)},  \label{53a} \\ \n \\
\rho + p &=& \frac{\alpha  \left(\frac{12 k_1 r^{2 k_1}}{c_1+r^{2 k_1}}-6 k_1+10\right)-4 \alpha  k+k-1}{k r^2}  \label{53b}, \\ \n \\
\rho + 3p &=& -\frac{2 \left(c_1 \left(10 \alpha -4 \alpha  k-6 \alpha  k_1+k+3 k_1-4\right)+\left(10 \alpha -4 \alpha  k+6 \alpha  k_1+k-3 k_1 -4\right) r^{2 k_1}\right)}{k r^2 \left(c_1+r^{2 k_1}\right)}.   \label{53c}
 \eeq
\end{widetext}
and each of these is expected to be positive. 

\begin{figure}
 \includegraphics[width=8cm]{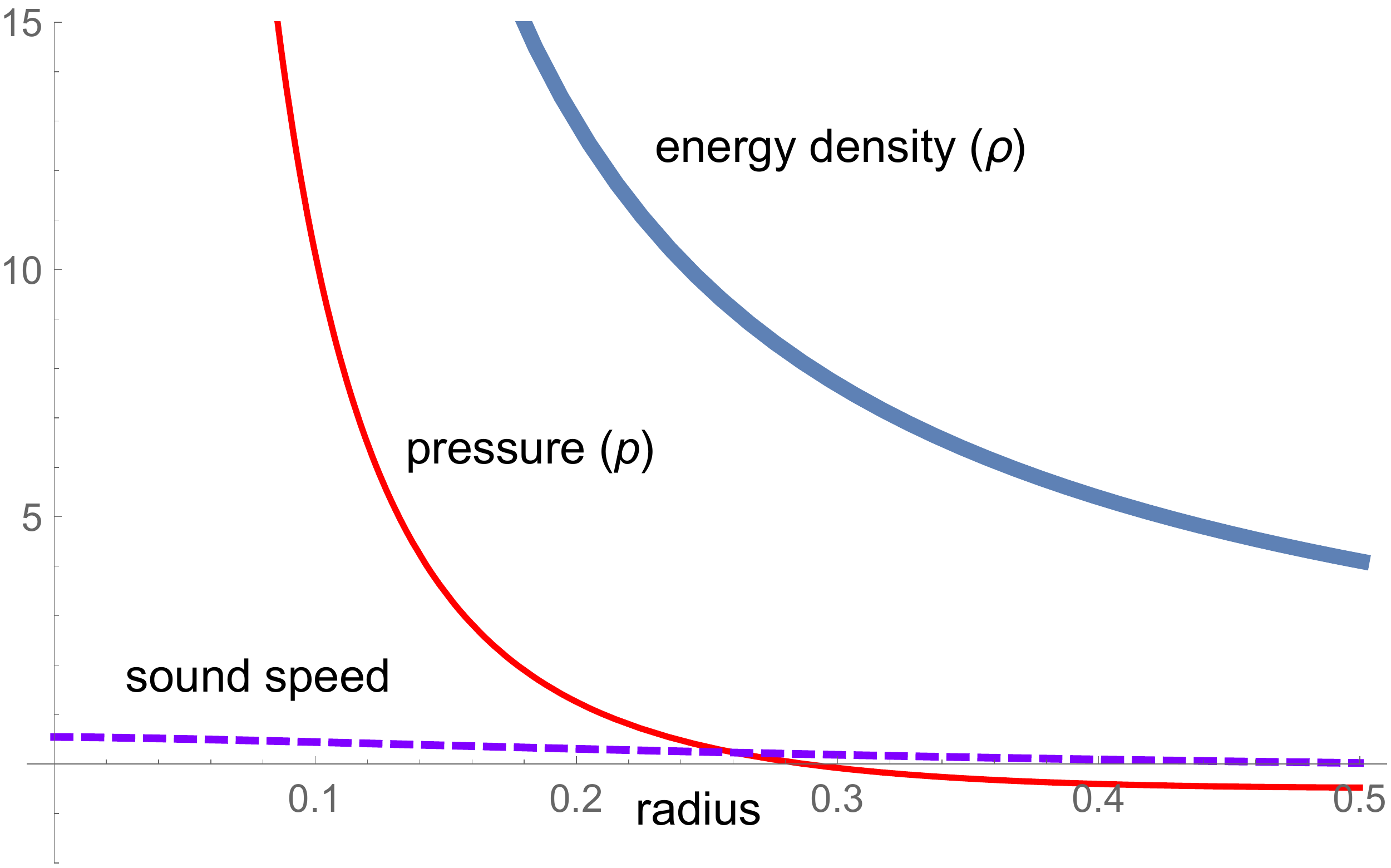}\\
  \caption{Dynamical quantities versus radial value $r$}\label{Fig 1}
\end{figure}

\begin{figure}
  \includegraphics[width=8cm]{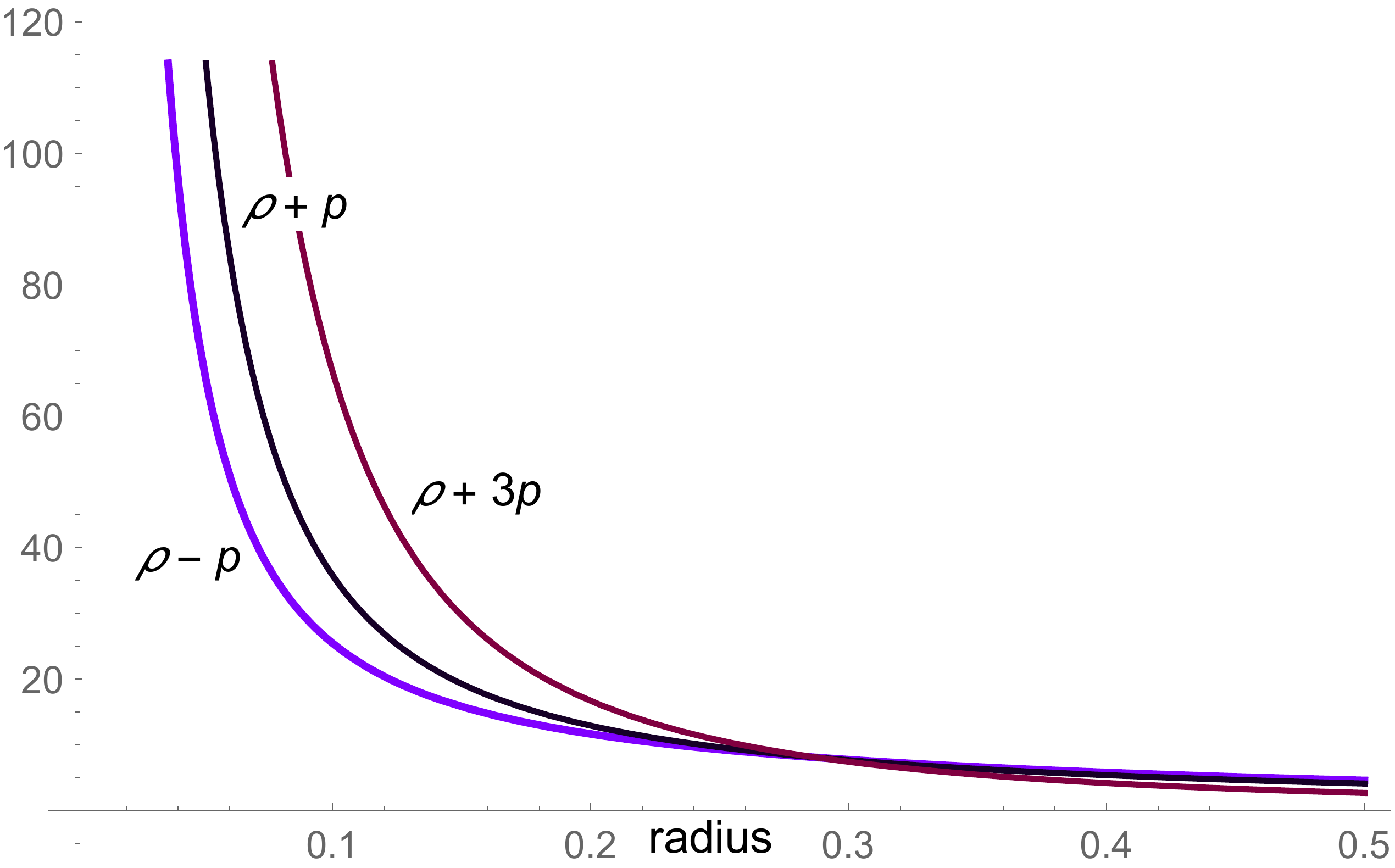}\\
  \caption{Energy conditions versus radial value $r$}\label{Fig 2}
\end{figure}
Finally, we provide some analysis of this model with the help of graphical representations for specific parameter values $\alpha = 0.5$, $c_1 = 1$ and $k = 1.5$. These values were selected after a process of fine-tuning to detect suitable parameter values.  Fig. (\ref{Fig 1}) demonstrates that the density and pressure are positive  with the pressure vanishing for the radial value $r = 0.287524$. Within this radius the sound speed is subluminal with $0 < \frac{dp}{d\rho} < 1$. Fig. (\ref{Fig 2}) depicts the energy conditions and it can be seen that the weak, strong and dominant energy conditions are well behaved within the proposed radii for this model. There are no asymptotes or singularities within the radius so the model is regular everywhere.  Accordingly we may conclude that this model satisfies all the elementary physical requirements for plausibility and is eligible to represent a realistic stellar distribution.  Noteworthy is also the fact that isothermal behaviour is absent since the pressure does not fall-off as $\frac{1}{r^2}$ like the energy density.

\section{Linear barotropic EoS}
\label{barotropic}

In this section, we investigate the consequences of stipulating the EoS $p = \epsilon \rho$ with range $\epsilon \in (0, 1)$. Again the problem is completely determined and a unique solution may in theory exist. With this linear barotropic  EoS, Eqs. (\ref{7a}) and (\ref{7b}) allow us to write the temporal potential in terms of the spatial potential in the following way: 
\begin{eqnarray} \nu' =
\frac{(1+\epsilon)(4\alpha-1)(e^{\lambda} -1) + (\alpha -
\epsilon(\alpha - 1))r\lambda'}{((3\alpha -1) + 3\alpha \epsilon)r}. \n \\ \label{60}
\end{eqnarray}
Now  plugging (\ref{60}) into (\ref{6c}) gives the relationship
\beq b^2 \left(a_1^2+a_1 r (2 a_2+a_3) b' + 4 a_1 a_3 - 4
a_3^2\right)   \n  && \\ \n \\ 
-2 \left(a_1^2+2 a_1 a_3-2 a_3^2\right) b^3 +a_1^2 b^4  \n &&  \\
\n \\ +rb \left((a_1 (a_3-2 a_2)-2 a_3 (a_2+a_3)) b'
+2 a_2
a_3 r b''\right) \n  && \\ \n \\ 
+a_2 r^2 (a_2-3 a_3) b'^2 &&= 0 ,\n \\ \label{62} \eeq
where we have defined $a_1 = (1 + \epsilon)(4\alpha - 1)$, $a_2 =
\alpha + \epsilon(\alpha -1)$, $a_3 = (3\alpha -1)+3\epsilon \alpha$ and
$e^{\lambda} = b(r)$.

The form of the nonlinear differential equation (\ref{62}) prohibits
locating the general solution for all values of the constants.
Accordingly in our pursuit for an exact solution, we consider
various values of the constants and relationships between them.

\subsection{$a_2 = 0$}

Setting $a_2 = 0$ relates the EoS proportionality constant
$\epsilon$ and the Rastall parameter $\alpha$ through $\epsilon =
\frac{\alpha}{\alpha -1}$. Eq (\ref{62}) simplifies to
\beq (b-1) b
\left(a_1^2 b-a_1^2-4 a_1 a_3+4 a_3^2\right) && \n \\
+a_3 r b' (a_1 b+a_1-2
a_3) && = 0, \label{63} \eeq 
 which may be rearranged in the form
\be
\frac{b'(b + k_1)}{b(b-1)(b-k_2)} = -\frac{a_1}{a_3 r}, \label{64}
\ee
where $k_1 = \frac{a_1-2a_3}{a_1}$ and $k_2 = \frac{a_1^2 + 4a_1
a_3 -4a_3^2}{a_1^2}$.

Eq. (\ref{64}) integrates to yield the implicit solution
\be
\frac{b^{k_1(k_2 -1)} (b-k_2)^{k_1 +k_2}}{(b-1)^{k_2(k_1 +1)}} =
Cr^{-\frac{a_1 k_2 (k_2 -1)}{a_3}}, \label{65}
\ee
where $C$ is an integration constant. In this form the solution is not useful. We seek values for $k_1$ and $k_2$ that will allow us to solve explicitly for $b$ in terms of $r$. If we now introduce the relationship $a_1 = 2a_3$,  then $k_1 = 0$ and $k_2 = 8$. Eq. (\ref{65}) assumes the simpler form
\be
\frac{b-8}{b-1} = C_1 r^{-14},  \label{66}
\ee
where $C_1 = C^{-14}$. The exact solution can now be expressed as 
\be
b(r) = e^{\lambda} = \frac{8r^{14} -C_1}{r^{14} - C_1}, \label{67} 
\ee
whereas the temporal potential evaluates to 
\be
e^{\nu} = C_2 r^{2(C_1  -1)} \left(1 - r^{14}\right)^{\frac{8-C_1}{7}}, \label{68}
\ee
with the aid of (\ref{8b}). Although we have succeeded in generating an exact solution with the $p = \epsilon \rho$, we note that the restrictions introduced result in the undesirable values $\epsilon = -\frac{1}{5}$ and $\alpha = -\frac{1}{4}$. The negative Rastall parameter $\alpha$ may be tolerated however the negative $\epsilon$ indicates a violation of causality as the sound speed index  now is $\frac{dp}{d\rho} = -\frac{1}{4}$. In the literature there exists works demonstrating that ultrabaric matter could be superluminal \cite{caporaso, glass}. Additionally Kistler {\it et al} considered cold neutron stars and concluded that these objects become superluminal and ultrabaric at densities of around $10^{15} g/cm^3$ \cite{kistler}. 

\subsection{$k_2 = 0$} 

Setting $k_2 = 0$ generates the potential $b = e^{\lambda} =$ a constant. This case has already been dealt with earlier. 

\subsection{$k_1 = 0$}

Prescribing $k_1 = 0$ implies $a_1 = 2a_3$ consequently $k_2 = 2$  and in turn this generates the relationship $\epsilon = \frac{1-2\alpha}{1+2\alpha}$. Requiring that the sound speed remains subluminal yields the constraint $0 < \alpha < \frac{1}{2}$ which is reasonable. With these assumptions Eq (\ref{65}) reduces to
\be
\frac{b-k_2}{b-1} = Cr^{-\frac{a_1 (k_2 -1)}{a^3}}, \label{69}
\ee
or 
\be
b(r) = e^{\lambda} = \frac{C r^{\frac{a_1 (k_2 -1)}{a_3}}-k_2}{C r^{\frac{a_1 (k_2 -1)}{a_3}}-1}. \label{70}
\ee
The remaining gravitational potential has the form 
\be
e^{\nu} = \frac{C_3\left(r^{2k_2} -Cr^2 \right)^{1-\frac{a_2}{a_3}} \left(k_2 r^{2k_2} -Cr^2 \right)^{\frac{a_2}{a_3}}}{r^2}, \label{71} 
\ee
where $C_3$ is a second integration constant. Note that the values of the constants may be given in terms of the Rastall parameter $\alpha$ as $a_1 = \frac{2(4\alpha -1)}{1+2\alpha}$, $a_2 =  a_3 = \frac{4\alpha -1}{1+2\alpha}$. The pressure and density evaluate to
\be
p = \epsilon \rho = \frac{(2\alpha -1)(3C-2r^2)}{(C -2r^2)^2}, \label{72} 
\ee
for the case $k_1 = 0$. Observe that a surface of vanishing pressure exists at $r = \sqrt{\frac{3C}{2}}$ thus a compact stellar model conforming to the elementary physical requirements has been successfully constructed. The negative aspect is the presence of an asymptote at $r = \sqrt{\frac{C}{2}}$ and the causality criterion $\alpha < \frac{1}{2}$ forces the interval of validity to be $r > \sqrt{\frac{3C}{2}}$ for a positive pressure profile.  
To circumvent this and eliminate the asymptote the value of $C$ should satisfy $C < 0$. Unfortunately this has the consequence that the surface of vanishing pressure is forfeited. However,  this is not physically unreasonable as the model may be interpreted  as representing a cosmological fluid with a linear barotropic equation of state. 

\section{Conclusion}
\label{Con}

In this article, we have investigated the role of a linear barotropic EoS on stellar structure within the framework of Rastall theory of gravity. The field equations were solved for the isothermal fluid prescriptions of an inverse square fall off of the density as well as a linear EoS. A physically reasonable model was generated depending on the value of the Rastall parameter.  In contrast the Einstein model published in the literature is known to be defective. When the EoS was relaxed an exact model was found for the parameter $\alpha = 1$. However, it was demonstrated that the pressure was inherently negative and consequently the energy conditions failed and the fluid was also acausal.  The configuration of a constant spatial gravitational potential was examined and a coherent model emerged. For suitable choices of parameters, a subluminal sound speed, positive energy density and pressure were guaranteed. Additionally a surface of vanishing pressure existed thus defining the boundary of the star. All the  energy conditions were satisfied. Finally, the case of an EoS was discussed. While the general exact solution could not be determined, it was found that solvable models existed for certain combinations of constants. Two classes of exact solutions were reported however they displayed non-physical properties. This study has shown that the Rastall framework is viable as a theory of gravity as models consistent with physical reality were generated. In some cases, like the isothermal cse, the Rastall model outshone its Einstein counterpart in that all the elementary physical requirements were met. 

\section{Acknowledgments}
SH thanks the National Research Foundation of South Africa for support under the competitive programme for rated researchers. AB thanks the University of UKZN for support.

%\appendix
%\section{}

%\begin{acknowledgments}
%\end{acknowledgments}

\bibliography{basename of .bib file}

\end{document}